\documentclass[iop,apj]{emulateapj}

\usepackage{graphicx}
\usepackage{amssymb}
\usepackage{amsmath}

\shorttitle{Approximations of rays in Schwarzschild field}
\shortauthors{O. Semer\'ak}

\begin{document}

\title{Approximating light rays in the Schwarzschild field}

\author{O. Semer\'ak}
\affil{Institute of Theoretical Physics,
       Faculty of Mathematics and Physics,
       Charles University in Prague, Czech Republic}
\email{oldrich.semerak@mff.cuni.cz}

\begin{abstract}
A short formula is suggested which approximates photon trajectories in the Schwarzschild field better than other simple prescriptions from the literature. We compare it with various ``low-order competitors", namely with those following from exact formulas for small $M$, with one of the results based on pseudo-Newtonian potentials, with a suitably adjusted hyperbola, and with the effective and often employed approximation by Beloborodov. Our main concern is the shape of the photon trajectories at finite radii, yet asymptotic behaviour is also discussed, important for lensing. An example is attached indicating that the newly suggested approximation is usable---and very accurate---for practically solving the ray-deflection exercise.
\end{abstract}


\section{Introduction}

Geodesic motion in a Schwarzschild field is one of introductory exercises in general relativity. The motion being completely integrable and planar (usually laid at $\theta=\pi/2$), its exact solution is mostly expressed as a $\phi(r)$ dependence. However, this involves elliptic integrals \citep{Hagihara-31,Darwin-59,MielnikP-62,Chandrasekhar-83} and may be quite uncomfortable if one needs to invert the result for $r(\phi)$ or for some parameter (usually the impact parameter). The solution for radius (or its reciprocal) as a function of angle was only given later
\citep{Rodriguez-87,KraniotisW-03,HackmannL-08,Scharf-11,Kostic-12,GibbonsV-12} using elliptic functions.
In particular, null geodesics (photon world-lines) have been treated notably by \citet{CadezK-05} and \cite{Munoz-14}.

However, a {\em simple and easily invertible} approximation of the relativistic photon trajectories seems yet to be suggested. Although light bending has been treated at many places, usually the {\em total bending angle} is the aim (given by directions at the source and observer locations, or just by radial asymptotics), especially when the exercise is treated in connection with gravitational lensing --- see e.g. \cite{VirbhadraE-00}, \cite{MutkaM-02}, \cite{Amore-etal-07}, \cite{ConnellF-08}, \cite{Virbhadra-09}, or \cite{Bozza-10} (recently the discussion has been mainly focused on the effect of the cosmological constant, e.g. \citealt{BhadraBS-10}, \citealt{BiressaF-11}, and \citealt{ArakidaK-12}). Instead, we would like to reasonably approximate {\em whole trajectories}, which is of course more delicate. (An even higher level would also incorporate proper {\em timing}, which is also important, but we restrict this study to spatial trajectories.) Actually, one learns quickly that formulas obtained by linearization in some parameter do not reproduce well the strong-field behaviour, while, on the contrary, ``improving" this by hand tends to spoil their weak-field limit. Low-order prescriptions typically do not provide sufficient bending in the centre's vicinity and sufficiently quick straightening at larger distances, so even if they may be adjusted to have a correct pericentre radius as well as asymptotic directions, their overall shape is often far from satisfactory.

Below, we first recall basic equations and fix parameterization of the problem (section \ref{Schwarzschild}). In section \ref{approximation}, several simple approximations of light rays, resulting from quite different approaches, are listed and their basic properties reviewed. Their performance at different radii is illustrated then and compared numerically with that of our simple suggestion presented (section \ref{examples}), showing that the latter is very accurate, even including trajectories with pericentres slightly below $r=4M$. Although we primarily focus on the behaviour of the approximations at finite radii from the central black hole, section \ref{asymptotic-angle} shows what answers the best of them give for an asymptotic angle along which the photons approach radial infinity. In section \ref{lensing-exercise} we illustrate the practical usage of our formula for solving the ray-deflection exercise, again with results almost identical to those obtained (purely numerically) from an exact treatment. Observations made are then briefly summarized in concluding remarks.

\section{Null geodesics in the Schwarzschild space-time}
\label{Schwarzschild}

Mainly to fix notation, let us recall basic equations of the exact problem.
Using Schwarzschild coordinates $(t,r,\theta,\phi)$ and geometrized units (in which $c=1$, $G=1$), we consider metric in the standard form
\[{\rm d}s^2=-N^2{\rm d}t^2+\frac{{\rm d}r^2}{N^2}+r^2({\rm d}\theta^2+\sin^2\theta\,{\rm d}\phi^2)\]
with $N^2:=1-2M/r$.
Geodesic motion in the spherically symmetric field being planar, let us choose the orbital plane to be the equatorial one ($\theta=\pi/2$). In such a case, the photon four-momentum has non-zero components
\begin{equation}  \label{photon-momentum}
  p^t=\frac{E}{N^2} \,, \quad
  p^r=\epsilon^r\,\frac{E}{r}\,\sqrt{r^2-N^2 b^2} \;, \quad
  p^\phi=\frac{L}{r^2} \,,
\end{equation}
where $E:=-p_t$, $L:=p_\phi$ and $b:=|L|/E$ are the photon's energy, angular momentum and impact parameter, respectively, all remaining conserved along the ray; the sign $\epsilon^r\equiv\pm 1$ fixes the orientation of radial motion (while that of azimuthal motion is determined by the sign of $L$).

Let us focus on photons with $b>3\sqrt{3}\,M$ which have a (one) turning point of radial motion, either pericentre (which is always above $r=3M$) or apocentre (which is always below $r=3M$).\footnote
{Photons with $b<3\sqrt{3}\,M$ move from infinity to the centre or vice versa without any radial turning.}
Let us then adjust, without loss of generality, the coordinates so that a given photon reaches this turning point at $\phi=0$. The ray thus gets symmetric with respect to the meridional plane $\{\phi=0,\pi\}$ and it is sufficient to only consider its half starting from that plane. The vanishing of radial momentum at $\phi=0$ constrains the constants of motion by the condition
\begin{equation}  \label{EL-constraint}
  r_0^2-N_0^2 b^2=0 \quad \Longrightarrow \quad
  b=\frac{r_0}{\sqrt{1-\frac{2M}{r_0}}} \;,
\end{equation}
where $r_0:=r(\phi=0)$ indicates the radius at the turning point of radial motion and $N_0:=N(r=r_0)$.

Equations for $p^\phi$ and $p^r$ give
\begin{equation}  \label{dphi/dr}
  \frac{{\rm d}\phi}{{\rm d}r}
  =\frac{\epsilon^r}{r}\,\frac{1}{\sqrt{\frac{r^2}{b^2}-N^2}}
\end{equation}
which can further be expressed in terms of the extremal radius $r_0$,
\begin{align}  \label{dphi/dr,r0}
  \frac{{\rm d}\phi}{{\rm d}r}
  &= \frac{\epsilon^r\,\frac{r_0}{r}}
          {\sqrt{N_0^2 r^2-N^2 r_0^2}} \nonumber \\
  &= \frac{\epsilon^r\,r_0^{3/2}}
          {\sqrt{r(r-r_0)}\;
           \sqrt{r(r+r_0)(r_0-2M)-2Mr_0^2}} \;.
\end{align}
This second form is more complicated, but it turns out to be much more suitable for integration.
Assuming, without loss of generality, that we focus on the half of the trajectory which starts toward positive $\phi$ from $\phi=0$ (briefly, we assume $L>0$), the integration gives (see \citet{Darwin-59} or \citet{Chandrasekhar-83}, formula (260) in Chapter 3)
\begin{align}
  \phi(r)
  &= \frac{2\,\sqrt{r_0}}{[(r_0-2M)(r_0+6M)]^{1/4}}
     \left[K(k)-F(\chi,k)\right]  \label{phi(r)} \\
  &= \frac{2\,\sqrt{r_0}}{[(r_0-2M)(r_0+6M)]^{1/4}}\;
     F(\chi',k) \,, \label{phi(r)'}
\end{align}
where $F(\chi,k):=\int_0^\chi\frac{{\rm d}\alpha}{\sqrt{1-k^2\sin^2\alpha}}$
is the elliptic integral of the 1st kind, with amplitude $\chi$ and modulus $k$ given by
\begin{align}
  \sin^2\chi &:=
    1-\frac{1}{k^2}\,
      \frac{2M\left(1-\frac{r_0}{r}\right)}{\sqrt{(r_0-2M)(r_0+6M)}}
    \;, \\
  2k^2 &:= 1-\frac{r_0-6M}{\sqrt{(r_0-2M)(r_0+6M)}} \;,  \label{2k2}
\end{align}
and $K(k):=F(\pi/2,k)$ is its complete version.
One can check immediately that $F(\chi,k)$ only reduces to $K(k)$ at the turning point, where $r=r_0$ and so $\chi=\pi/2$, which correctly yields $\phi(r\!=\!r_0)=0$.
The second expression (\ref{phi(r)'}) contains a different amplitude $\chi'$ which is related to $\chi$ by
\begin{align}
  \sin^2\chi' &= \frac{1-\sin^2\chi}{1-k^2\sin^2\chi} \nonumber \\
              &= \frac{4Mk^{-2}\left(1-\frac{r_0}{r}\right)}
                      {\sqrt{(r_0-2M)(r_0+6M)}+r_0-2M-4M\frac{r_0}{r}}
\end{align}
(while $k$ remains unprimed in both expressions).
We add that the complementary modulus $k'$, which is related to $k$ by $k'^2=1-k^2$, is given by the same expression (\ref{2k2}) as $k$, just with {\em plus} after 1; their product is therefore quite short,
\begin{equation}
  k^2 k'^2=\frac{4M\,(r_0-3M)}{(r_0-2M)(r_0+6M)} \;.
\end{equation}

With our parametrization, the azimuth $\phi$ of a photon increases monotonically from zero. For a photon starting (from $\phi=0$) from $r_0\simeq 3M$ the azimuth can finally reach very large values; with $r_0$ growing from $3M$ to infinity the asymptotic azimuth decreases from infinity to $\pi/2$, while with $r_0$ shrinking from $r=3M$ toward $2M$ the photon falls through the horizon at $\phi$ quickly decreasing from infinity toward zero.

\section{Approximating the light rays}
\label{approximation}

The exact Darwin's formula (\ref{phi(r)}) is quite simple, yet it is not easy to invert it for $r(\phi)$ and mainly for $r_0$ (which would in turn yield constants of the motion as functions of $r$ and $\phi$). Such a problem is typically encountered when asking ``What are the parameters of the light/photons that arrive at a given location from some (which?) points of a given source?" (in the Schwarzschild field). Namely, the source generally emits photons of various different parameters (from different starting points, with different energies, in different directions, etc.), of which each follows a different world-line. When studying some radiation-influenced process at a given location outside the source, it is first necessary to find {\em which} of the photons get to that location (which means to find from where in the source they started) and then to infer what are their properties (in order to be able to say what will be their effect).

An important system that raises such questions is an accretion disc around a black hole, because i) its inner part is a powerful source of radiation which strongly affects matter around (on the other hand, the disc can be significantly irradiated by a hot ``corona", and possibly even self-irradiated), and ii) the inner part of the disc lies in a very strong gravitational field where the propagation of light is not trivial (linear).
To give a specific example, we came across the demand to invert Darwin's formula for $r_0$ when studying, in Schwarzschild space-time, the motion of test particles influenced by a radiation field emitted from the equatorial plane in a perpendicular direction (such a pattern was considered to approximate the radiation generated by an equatorial thin disc) --- see \cite{BiniGJS-15}. In such an arrangement, one needs to find the impact parameters of the (two) ``vertical" photons interacting with the particle at a given location, which precisely requires one to find the $r_0$'s from where those photons started.

Therefore, it is worthwhile to look for an approximation of photon trajectories which would be simple, invertible for $r_0$, yet reasonably accurate. In particular, it should work well for as small a radius ($r_0$) as possible, because in real astrophysical situations there is often much radiation just quite close to the horizon. This is natural since matter inflows into these regions with extremely high speeds, so collisions of its streams dissipate huge amounts of energy which intensively outflow as radiation. In particular, real accretion discs are supposed to radiate most from their innermost regions which probably reach to the innermost stable circular geodesic at $r=6M$ or even below.

A natural starting attempt for how to simplify the inversion problem is to linearize the exact formula in some small parameter. We have either $2M<r\leq r_0<3M$ or $r\geq r_0>3M$, with $b>r_0$ anyway. Clearly the $r_0>3M$ case is more astrophysically interesting. Then the relation between $r$ and $b$ changes in time: the photon starts from $r\equiv r_0<b$, but quickly gets to $r>b$ (and then even to $r\gg b$). Surely $M$ is the least of all parameters and hence the usual linearization in it. However, the weak-field approximations --- like the one obtained by linearization in $M$ --- generally yield trajectories ``less bent about" the central gravitating body, since the centre's field is weakened effectively. If such an approximation is employed for the inversion of the $\phi(r;r_0)$ formula for $r_0$, it may lead to errors when applied to strongly bent rays. Actually: adopt our parametrization, i.e., adjust the plane $\phi=0$ so that the ray crosses it (``starts from it") in a perpendicular direction (it is purely azimuthal there). Now, imagine a photon approaching $\phi=\pi$ (from $\phi<\pi$) while having a small radius: it must have started (purely tangentially) from $\phi=0$ from a {\em very small} radius $r_0$ in order for it to have been bent about the centre sufficiently. For such photons, weak-field approximations may easily yield a starting radius $r_0$ even lying {\em below horizon} (note that there is actually no horizon in most such approximations), which may then bring errors if substituted into the Schwarzschild-metric lapse function $N\equiv\sqrt{1-2M/r}$.

One could surely improve the approximation by taking into account higher order(s) of the small-parameter expansion, but higher-order formulas mostly can{\em not} be inverted easily. Restricting oneself to low-order formulas, one can resort to some more ``pragmatic" construct instead. First, ``pseudo-Newtonian" descriptions are often employed in the astrophysical literature, based on the Newtonian potential suitably modified to mimic certain features of the actual relativistic field. Second, our particular problem of scattering-type light trajectories can be approximated by a hyperbola adjusted to the desired asymptotic directions. And third, one can try to design a suitable formula ``by hand", simply observing the main properties it should reproduce. However, all such ad hoc formulas, not relying on any clearly justified procedure, must be handled with care; in particular, even if they were successful close to the black hole (or rather {\em more if} they were successful there), their weak-field (large-radius) behaviour may not be satisfactory. They can also hardly yield trustful replies for tiny, velocity-dependent (dragging), non-stationary, radiation etc. effects, but in the simple case of static space-times, they have mostly proved quite practical. Let us compare the above possibilities, including mainly several successful suggestions from literature.

\begin{figure*}
\epsscale{.85}
\plotone{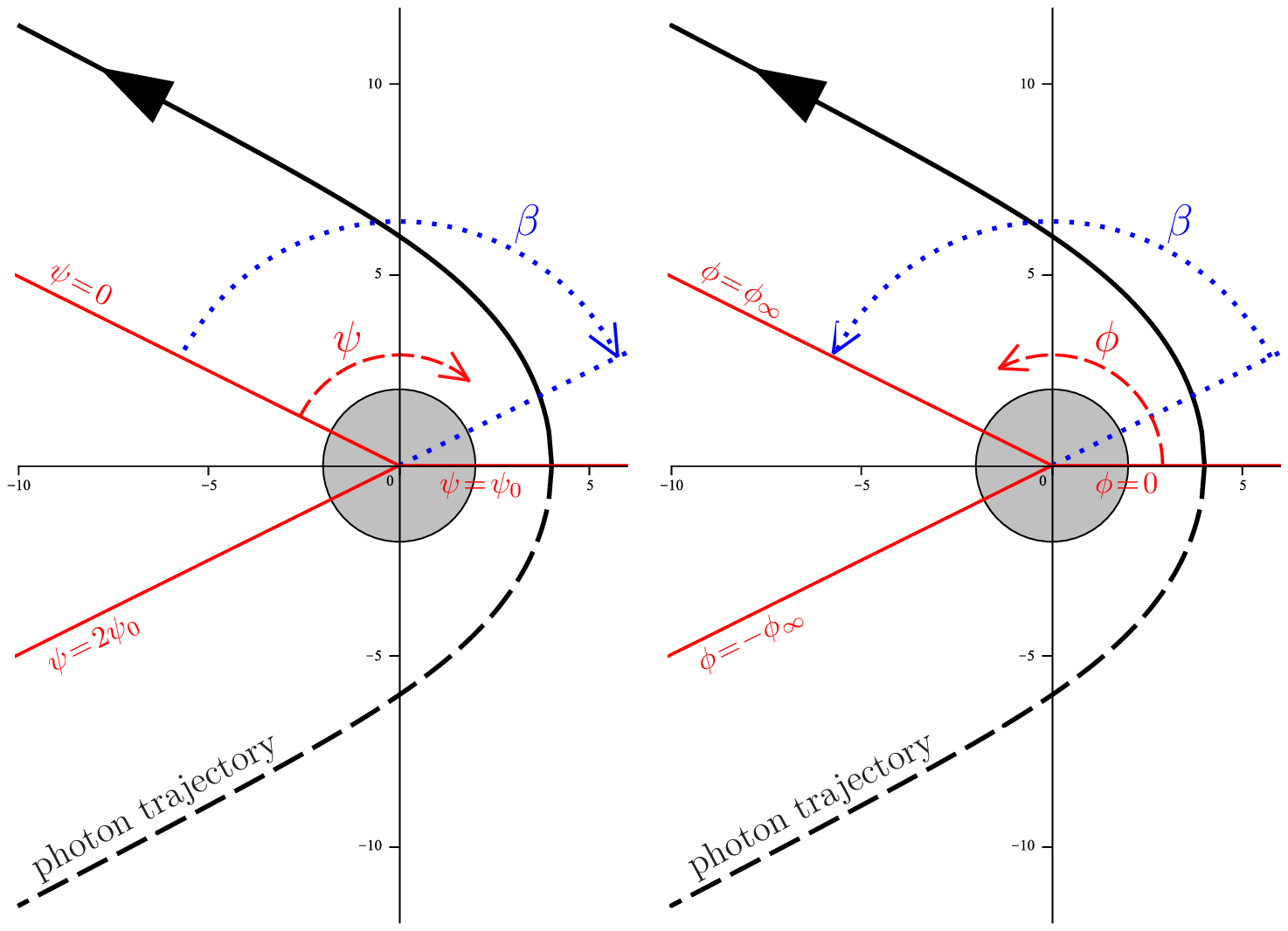}
\caption
{Azimuthal description of a photon trajectory used by Beloborodov \citep{Beloborodov-02} ($\psi$, {\it left}) and by us in this note ($\phi$, {\em right}); $\beta$ is the total deflection angle. Axes $r\cos\phi$, $r\sin\phi$ are also shown (with values in units of $M$), the black hole is gray.}
\label{ray-description}
\end{figure*}

\subsection{Linearization of Darwin's formula in $M$}
\label{linearization-in-M}

Darwin's formula (\ref{phi(r)}) can be linearized in $M$ to obtain
\begin{equation}  \label{Darwin,lin}
  \cos\phi=
  \frac{r_0}{r}-\frac{M}{r_0}\,\frac{(r-r_0)(2r+r_0)}{r^2}
  +O(M^2) \,.
\end{equation}
Clearly $\phi=0$ corresponds to $r=r_0$ correctly, while for $r\rightarrow\infty$ one has asymptotics $\cos\phi_\infty=-2M/r_0$ which is always $>-1$ (so $\phi_\infty<\pi$ which means that the ray's half-bending is less than $90^\circ$). The linear part can easily be inverted for $r_0$.

\subsection{First $M$-orders from the Binet formula}

Although the treatments on lensing mostly aim at the total deflection angle, some of them also provide prescription for the photon trajectory. Both are usually obtained by perturbative solution of the well-known Binet formula which in the null case reads
\begin{equation}
  \frac{{\rm d}^2 u}{{\rm d}\phi^2}+u=3Mu^2,
  \qquad {\rm where} \quad u:=\frac{1}{r} \,.
\end{equation}
Let us mention three of them.
\cite{BiressaF-11} solved the equation up to linear order by\footnote
{Papers on lensing usually adjust the azimuth so that the pericentre lies at $\phi=\pi/2$, so we must shift their formulas by $\pi/2$.}
\begin{align}
  \frac{r_0}{r} &= \cos\phi+\frac{M}{r_0}\,(1+\sin^2\phi-\cos\phi) \nonumber \\
                &= \cos\phi+\frac{M}{r_0}\,(2+\cos^2\phi)(1-\cos\phi)
  \label{BiressaF,lin}
\end{align}
in their equation (14). Clearly $r(\phi=0)=r_0$ as it should be and solving for $\cos\phi$ yields (\ref{Darwin,lin}) in linear order.

A slightly different linear-order solution was given by \cite{BhadraBS-10} (their equation (5)),
\begin{equation}
  \frac{R}{r}=\cos\phi+\frac{M}{R}\,(1+\sin^2\phi),
\end{equation}
where $R$ is a length whose meaning follows by setting $\phi=0$ (and $r=r_0$):
\begin{equation}
  \frac{R}{r_0}=1+\frac{M}{R} \,.
\end{equation}
Solving this for $R$, substituting above and linearizing in $M$, one has
\begin{equation}  \label{BhadraBS,lin}
  \frac{r_0+M}{r}=\cos\phi+\frac{M}{r_0}\,(1+\sin^2\phi).
\end{equation}
Although this differs from (\ref{BiressaF,lin}), its solution for $\cos\phi$ again agrees with (\ref{Darwin,lin}) in linear order.

Finally, let us turn to \cite{ArakidaK-12} who presented a second-order solution in their equation (7),
\begin{align}
  \frac{b}{r}
  =\cos\phi
   &+\frac{M}{b}\,(1+\sin^2\phi) \nonumber \\
   &+\frac{M^2}{4b^2}\,(7\cos\phi+15\,\phi\,\sin\phi+3\cos^3\phi) \,.
\end{align}
The corresponding equation expressed in terms of $r_0$ follows by substituting (\ref{EL-constraint}) for $b$ and expanding in $M$ accordingly,
\begin{align}
  & \frac{r_0+M}{r}+\frac{3M^2}{2r_0 r}
    =\cos\phi+\frac{M}{r_0}\,(1+\sin^2\phi) \nonumber \\
  & +\frac{M^2}{4r_0^2}\,(7\cos\phi+15\,\phi\,\sin\phi+3\cos^3\phi-4-4\sin^2\phi) \,.
  \label{ArakidaK,lin}
\end{align}
In linear order this reduces to (\ref{BhadraBS,lin}) and its $\phi=0$ form
\[\frac{r_0+M}{r}+\frac{3M^2}{2r_0 r}=1+\frac{M}{r_0}+\frac{3M^2}{2r_0^2}\]
properly yields $r=r_0$.
Inversion for $b$ or $r_0$ is of course more uncomfortable if keeping the second order in $M$.

\subsection{Using a suitable pseudo-Newtonian potential}
\label{pseudo}

In Newton's theory, the motion of test particles in the velocity-independent spherical potential $V(r)$ is also confined to a plane (which we again identify with $\theta=\pi/2$) and described by
\begin{equation}  \label{ddot:r,phi}
  \ddot{r}=-V_{,r}+r\dot{\phi}^2,
  \qquad
  r\ddot{\phi}=-2\dot{\phi}\,\dot{r} \,.
\end{equation}
These equations have usual integrals of energy and angular momentum
\begin{equation}
  E=\frac{m}{2}\,(\dot{r}^2+r^2\dot{\phi}^2)+mV,
  \qquad
  L=m r^2\dot{\phi} \,,
\end{equation}
which invert for velocities as
\begin{equation}  \label{velocities}
  r\dot{\phi}=\frac{L}{mr} \,,
  \qquad
  \dot{r}^2=\frac{2mr^2(E-mV)-L^2}{m^2 r^2} \;.
\end{equation}
The ratio of the velocities gives an equivalent of the relativistic equation (\ref{dphi/dr}),
\begin{equation}  \label{dphi/dr,pseudo}
  \frac{{\rm d}\phi}{{\rm d}r}=
  \frac{\epsilon^r}{r}\left[\frac{2mr^2}{L^2}(E-mV)-1\right]^{-1/2},
\end{equation}
where $\epsilon^r=+1$ will again be chosen below.
Should the trajectory be strictly tangential (azimuthal) at $\phi=0$ ($\dot{r}=0$ at $r=r_0$), it has to satisfy the condition
\begin{equation}
  2mr_0^2(E-mV_0)=L^2 \,, \quad {\rm where} \;\;
  V_0:=V(r=r_0).
\end{equation}

Besides the overall caution in following the pseudo-Newtonian approach, it is also necessary to distinguish between various potentials proposed in the literature, because different ones are suitable for different purposes (see e.g. \citealt{Crispino-etal-11}).
The Paczy\'nski-Wiita potential $V=-M/(r-2M)$ is a most simple and efficient mimicker of the Schwarzschild field, whose main advantages are correct values of important circular geodesics, so it is mainly suitable for modeling accretion discs. However, if plugged into the above, the resulting equation for ${\rm d}\phi/{\rm d}r$ is not easier to integrate than its relativistic counter-part. The same also applies to the Nowak-Wagoner quadratic-expansion potential
\[V=-\frac{M}{r}\left(1-\frac{3M}{r}+\frac{12M^2}{r^2}\right)\]
which has often proved the best of the ``simple" suggestions.\footnote
{A more advanced yet elegant possibility, namely a potential suitably dependent on velocity, was suggested by \citet{TejedaR-13}. It would be worth checking whether it is also useful for null geodesics.}
It is clear from equation (\ref{dphi/dr,pseudo}) that for its integration to be simpler than that of the relativistic case (\ref{dphi/dr}), the resulting polynomial under the square root has to be exactly quadratic in $r$. This is the case if the potential is of the form
\begin{equation}  \label{potential-Wegg}
  V=-\frac{M}{r}\left(1+\frac{\alpha M}{r}\right),
\end{equation}
where $\alpha$ is some constant. Such a form has been advocated by \citet{Wegg-12}, specifically with $\alpha=3$. Equation (\ref{dphi/dr,pseudo}) supplied with the initial condition $\phi_0:=\phi(\dot{r}=0)=0$ is then (with the above form of the potential) solved by
\begin{equation}
  \phi(r)=
  \frac{1}{k}\;
  {\rm arccot}\frac{k-\frac{m^2 Mr}{kL^2}}
                   {\sqrt{\frac{2mr^2}{L^2}(E-mV)-1}} \;,
\end{equation}
where we introduced the dimensionless constant
\[k^2:=1-\frac{2m^2 \alpha M^2}{L^2} \;.\]

The pseudo-Newtonian picture is doubly problematic when describing photons, in particular, their speed cannot be considered constant. However, choosing their {\em initial} linear speed to be equal to one, which means $r_0\dot\phi_0=1$ in our case when photons start in a pure azimuthal direction, one has
\[E=\frac{m}{2}+mV_0, \quad
  L=m r_0, \quad
  k^2=1-\frac{2\alpha M^2}{r_0^2} \,,\]
so\footnote
{Note that the photon has enough energy to escape to infinity, $E>0$, if $V_0>-1/2$, which holds for $r_0\!>\!(1\!+\!\sqrt{1+2\alpha})M$; for $\alpha\!=\!3$ this means $r_0>3.65M$, which is not so bad.}
the equatorial trajectory can be rewritten as
\begin{align}
  \phi(r)
  &= \frac{r_0}{\sqrt{r_0^2-2\alpha M^2}} \,\times \nonumber \\
  & \quad \times \; {\rm arccot}
                    \frac{r_0^2-2\alpha M^2-Mr}
                         {\sqrt{r_0^2-2\alpha M^2}\;\sqrt{r^2-r_0^2+2r^2(V_0-V)}} \;.
  \label{Wegg-trajectory}
\end{align}
Clearly $\phi(r=r_0)=0$ correctly and the asymptotic value at radial infinity amounts to
\begin{align}
  \phi_\infty
  &= \frac{r_0}{\sqrt{r_0^2-2\alpha M^2}} \,\times \nonumber \\
  & \quad \times \left(\pi - {\rm arccot}\frac{M}{\sqrt{r_0^2-2\alpha M^2}\;\sqrt{1+2V_0}}\right).
  \label{Wegg,phi_infty}
\end{align}

When speaking about asymptotics, it should be stressed that photons exist which remain bound on ``elliptic-type" orbits. Actually, recalling equations (\ref{velocities}), we see that $\dot{r}=0$ if $2mr^2(E-mV)=L^2$, which after substitution of our $E=m/2+mV_0$, $L=m r_0$ implies $r^2-r_0^2=2r^2(V-V_0)$, i.e., with $\alpha=3$ and expanded,
\begin{equation}
  (r-r_0)\left[(r_0^2-6M^2)(r+r_0)-2Mrr_0\right]=0.
\end{equation}
Besides the automatic zero at $r=r_0$, this also has a second root at
\[r=-r_0\,\frac{r_0^2-6M^2}{r_0^2-2Mr_0-6M^2} \;.\]
However, this root is only relevant in a narrow interval $r_0\in(2.8922,3.6458)M$ (it grows from $2M$ to infinity very fast within this range).

There is one case of particular interest within the above range of $r_0$:
using $L\equiv m r^2\dot{\phi}=m r_0$ back in the first of equations (\ref{ddot:r,phi}), we have
\begin{equation}
  r^3\ddot{r}=-r^3 V_{,r}+r_0^2
             =r_0^2-Mr-2\alpha M^2 \,,
\end{equation}
which specifically for $\alpha=3$ reads
\begin{align*}
  r^3\ddot{r} &= r_0^2-Mr_0-6M^2+M(r_0-r) \\
              &= (r_0-3M)(r_0+2M)+M(r_0-r)
\end{align*}
and thus implies that $\ddot{r}=0$ at $r=r_0=3M$. Therefore, the potential (\ref{potential-Wegg}) with $\alpha=3$ correctly reproduces the Schwarzschild circular photon geodesic (which indicates that it may be successful in simulating photon motion in the innermost region).

\subsection{Approximation by a hyperbola}
\label{hyperbola-approximation}

Another possibility is to approximate the photon trajectory by a suitable hyperbola. Placing its focus at $r=0$ and its vertex at ($\phi=0$, $r=r_0$), and prescribing some asymptotic azimuth $\phi_\infty$, it is given by the equation
\begin{equation}  \label{hyperbola}
  r\cos\phi = r_0+(r-r_0)\,\cos\phi_\infty \,.
\end{equation}
Now $\phi_\infty$ can be prescribed somehow, for example chosen according to the exact formula. However, since the latter makes the above expression uncomfortably long, let us instead illustrate it with $\phi_\infty$ provided by the Beloborodov formula (see next subsection), i.e. with $\cos\phi_\infty=-\frac{2M}{r_0-2M}$ (which proved quite accurate):
\begin{equation}
  r\cos\phi = r_0-2M\,\frac{r-r_0}{r_0-2M} \,.
\end{equation}
The limit possibility is to choose $\phi_\infty=\pi$ which yields the parabola
\[r\cos\phi = 2r_0-r.\]

\subsection{Beloborodov's approximation}

A simple approximation of photon trajectories has been provided by \citet{Beloborodov-02} in his formula (3),
\begin{align}
  r(\psi)&= \sqrt{M^2\,\frac{(1-\cos\psi)^2}{(1+\cos\psi)^2}+\frac{b^2}{\sin^2\psi}}
            -M\,\frac{1-\cos\psi}{1+\cos\psi} \nonumber \\
         &= \sqrt{M^2\,\tan^2\frac{\psi}{2}+\frac{b^2}{\sin^2\psi}}
            -M\,\tan\frac{\psi}{2} \;,
\end{align}
where the position angle $\psi$ is measured (from the centre) so that $\psi=0$ fixes the asymptotic escape direction (with all the orbit having $\psi>0$, so the photon is considered to move {\em against} the $\psi$ orientation).
The value of $\psi$ at pericentre ($\psi_0$) is determined by
\begin{equation}
  \frac{{\rm d}r}{{\rm d}\psi}=0
  \quad \Longrightarrow \quad
  \cos\psi=-\frac{2M}{r_0-2M}=:\cos\psi_0
\end{equation}
and lies between $\pi/2$ and $\pi$. The angle $\beta=2\psi_0-\pi$ represents the total bending angle; $2\psi_0$ is the asymptotic {\em ingoing} direction if the trajectory is extended to infinity in both directions (see figure \ref{ray-description}).
Beloborodov derived the above result from the elegant equation
\begin{equation}  \label{alpha-r,psi}
  1-\cos\alpha=(1-\cos\psi)\,N^2
\end{equation}
which pretty accurately approximates the relation between the photon's momentary radius $r$, position angle $\psi$ and the local direction of flight measured by the angular deflection from the radial direction in a local static frame, $\alpha$. (How to understand the success of this ``cosine relation" is explained in section 2 of \citealt{Beloborodov-02}.) The local direction $\alpha$ is given by the photon momentum,
\begin{align}
  \tan\alpha &= \frac{\sqrt{g_{\phi\phi}}\,|p^\phi|}{\sqrt{g_{rr}}\,|p^r|}
              = \frac{bN}{\sqrt{r^2-b^2 N^2}} \nonumber \\
  &\Rightarrow \quad
    \sin\alpha=\frac{bN}{r} \,, \;\;
    \cos\alpha=\sqrt{1-\frac{b^2 N^2}{r^2}} \,,
\end{align}
so one has, by solving equation (\ref{alpha-r,psi}) for $\cos\psi$ and then substituting for $\cos\alpha$,
\begin{equation}  \label{cos(psi)}
  \cos\psi=\frac{r\cos\alpha-2M}{r-2M}
          =\frac{\sqrt{r^2-b^2 N^2}-2M}{r-2M} \;.
\end{equation}
This increases monotonically from $-\frac{2M}{r_0-2M}\equiv\cos\psi_0$ at pericentre through zero to positive values and approaches unity at $r\rightarrow\infty$.

In our parametrization the photon has pericentre at $\phi=0$ and moves in a positive $\phi$ direction (to some asymptotic $\phi_\infty\equiv\psi_0$), and hence the angles are related by $\psi=\psi_0-\phi=\phi_\infty-\phi$ (see figure \ref{ray-description}), which means
\begin{align}
  \cos\psi&= \cos(\psi_0-\phi) = \cos\psi_0\cos\phi+\sin\psi_0\sin\phi \nonumber \\
  \quad   &= -\frac{2M}{r_0-2M}\,\cos\phi
             +\frac{\sqrt{r_0(r_0-4M)}}{r_0-2M}\,\sin\phi \,.
          \label{cos(psi),phi}
\end{align}
Hence, equating (\ref{cos(psi),phi})=(\ref{cos(psi)}) gives the implicit relation
\begin{equation}  \label{Beloborodov-approx}
  \frac{N_0^2}{N^2}\left(\!\sqrt{1\!-\!\frac{b^2 N^2}{r^2}}-\frac{2M}{r}\!\right)
  =\sqrt{1\!-\!\frac{4M}{r_0}}\,\sin\phi-\frac{2M}{r_0}\cos\phi
\end{equation}
for the $(r,\phi)$ trajectory.

The Beloborodov's formula is only applicable at $r_0>4M$ (it yields $\phi_\infty=\pi$ for $r_0=4M$), but it really provides a very good approximation almost all the way down to there. It is less suitable for finding $r_0$ as a function of $r$ and $\theta$, because the above relation yields for it an equation of the 16th (or at least the 8th) degree.

\subsection{New suggestion}
\label{new-suggestion}

The main purpose of this note is to suggest and test another ray-approximating formula,
\begin{equation}  \label{our-approximation}
  \cos\phi=\frac{r_0}{r}-
           \frac{M}{r_0-\alpha M}\,
           \frac{(r-r_0)(2r+r_0)}{(r-\omega M)^2} \;,
\end{equation}
where $\alpha$ and $\omega$ are real constants.
It correctly gives $\cos\phi=1$ at $r=r_0$, its radial asymptotics reads
\begin{equation}  \label{our-approximation,infty}
  \cos\phi_\infty=-\frac{2M}{r_0-\alpha M}
\end{equation}
and it can be inverted for $r_0$ as
\begin{equation}  \label{our-approximation,r0}
  r_0=\frac{{\cal R} +\sqrt{{\cal R}^2+4Mr{\cal A}{\cal B}}}{2{\cal A}} \;,
\end{equation}
where
\begin{align}
  {\cal R} &\equiv (r-\omega M)^2 (r\cos\phi+\alpha M)-Mr^2 \,, \nonumber \\
  {\cal A} &\equiv (r-\omega M)^2+Mr, \nonumber \\
  {\cal B} &\equiv 2r^2-(r-\omega M)^2\alpha\cos\phi \,. \nonumber
\end{align}
The formula works reasonably within a certain range of parameters $\alpha$ and $\omega$ and it is hard to say which particular combination is the best, because accuracy at small radii favours somewhat different values than accuracy farther away. We will specifically show that very good results are obtained with $\alpha=1.77$ and $\omega=1.45$, for example.

As in the case of Wegg's pseudo-Newtonian potential, ``elliptic-type" bound orbits do exist. Since
\[\frac{{\rm d}r}{{\rm d}\phi}=-\frac{\sin\phi}{{\rm d}(\cos\phi)/{\rm d}r} \;,\]
this would require either $\sin\phi=0$, or ${\rm d}(\cos\phi)/{\rm d}r\rightarrow\infty$.
The latter would only hold for $r_0=\alpha M$ or $r=\omega M$, of which none applies with our choice of $\alpha$ and $\omega$ (namely, we always have $r_0>\alpha M$ and $r>\omega M$). Hence, purely tangential motion can only happen at $\sin\phi=0$. This holds automatically at $r=r_0$ (where $\cos\phi=+1$), and it can also hold on the other side of the equatorial plane, $\cos\phi=-1$. Solving this equation for $r$, one finds that for our constants $\alpha=1.77$, $\omega=1.45$ the solution grows very quickly from $2.025M$ to infinity with $r_0$ increased from $2M$ to $(2+\alpha)M$ (this last value is valid generally and does not depend on $\omega$). In other words, all photons launched from $r_0>(2+\alpha)M$ escape to infinity and have $\phi_\infty<\pi$ there, consistent with the asymptotic formula (\ref{our-approximation,infty}).

\subsection{Comparison on expansions in $M$}
\label{expansions}

Before embarking on a numerical comparison of the approximations with the exact formula, let us further check their algebraic properties by expanding them in powers of $M$. To be more specific, let us thus expand $\cos\phi(r)$.
First, the cosine of the exact formula (\ref{phi(r)}) expands as
\begin{align}
  \cos\phi = &\,\frac{r_0}{r}
              -\frac{M}{r_0}\,\frac{(r-r_0)(2r+r_0)}{r^2} \nonumber \\
             &-\frac{M^2}{r_0^2}\,\frac{r-r_0}{r}\,\frac{{\cal D}}{4r^2}
              +O(M^3) \,,
\end{align}
where abbreviation has been used
\[{\cal D}:= 30r^2\sqrt{\frac{r+r_0}{r-r_0}}\;{\rm arccos}\sqrt{\frac{r+r_0}{2r}}
             -8r^2+9rr_0+5r_0^2 \,.\]

Now to the approximations.
Of the prescriptions obtained from perturbative solution of the Binet formula, we choose the result (\ref{BiressaF,lin}) by \cite{BiressaF-11}. When expressed in terms of $\cos\phi$, it expands as
\begin{align}
  \cos\phi = &\,\frac{r_0}{r}
              -\frac{M}{r_0}\,\frac{(r-r_0)(2r+r_0)}{r^2} \nonumber \\
             &-\frac{M^2}{r_0^2}\,\frac{(r-r_0)(2r+r_0)(r+2r_0)}{r^3}
              +O(M^3) \,.  \label{Biressa,expansion}
\end{align}
Performing the same expansion with the pseudo-Newtonian result (\ref{Wegg-trajectory}) involving Wegg's potential $V=-(M/r)(1+3M/r)$, one obtains
\begin{align}
  \cos\phi = &\,\frac{r_0}{r}
              -\frac{M}{r_0}\,\frac{r-r_0}{r} \nonumber \\
             &-\frac{M^2}{r_0^2}\,\frac{r-r_0+3\sqrt{r^2-r_0^2}\;{\rm arccos}\,\frac{r_0}{r}}{r}
              +O(M^3) \,.
\end{align}
The hyperbola (\ref{hyperbola}), if ``endowed with" the exact asymptotic angle (given by Darwin's formula), expands to
\begin{align}
  \cos\phi = &\,\frac{r_0}{r}
              -\frac{2M}{r_0}\,\frac{r-r_0}{r} \nonumber \\
             &-\frac{M^2}{r_0^2}\,\frac{(15\pi-16)(r-r_0)}{8r}
              +O(M^3) \,.  \label{hyperbola,expansion}
\end{align}
Next approximation is the one owing to Beloborodov's relation (\ref{Beloborodov-approx}). Solving the latter for $\cos\phi$ and expanding as above, we get
\begin{align}
  \cos\phi = &\,\frac{r_0}{r}
              -\frac{M}{r_0}\,\frac{(r-r_0)(2r+r_0)}{r^2} \nonumber \\
             &-\frac{M^2}{r_0^2}\,\frac{r-r_0}{r_0}\,
               \frac{{\cal B}}{2r^3}
              +O(M^3) \,,
\end{align}
where
\[{\cal B}:= 8r^3+4r^2 r_0+5rr_0^2+3r_0^3-4r(2r-r_0)\sqrt{r^2-r_0^2} \;.\]
Finally, our formula (\ref{our-approximation}) expands as
\begin{align}
  \cos\phi = &\,\frac{r_0}{r}
              -\frac{M}{r_0}\,\frac{(r-r_0)(2r+r_0)}{r^2} \nonumber \\
             &-\frac{M^2}{r_0^2}\,\frac{(r-r_0)(2r+r_0)(\alpha r+2\omega r_0)}{r^3}
              +O(M^3) \,.  \label{our,expansion}
\end{align}

As expected, the absolute term (corresponding to a straight line) is the same for all of the formulas. The linear terms are also common (and corresponding to the expansion of the exact formula), except for the pseudo-Newtonian formula and the hyperbola. The quadratic terms are somewhat longer, only in cases (\ref{Biressa,expansion}), (\ref{hyperbola,expansion}) and (\ref{our,expansion}) they remain rather simple. In particular, notice that the expansion of our suggestion (\ref{our,expansion}) is a generalization of the expansion (\ref{Biressa,expansion}) obtained from the approximate solution by \cite{BiressaF-11}. It is also seen that if our constants $\alpha$ and $\omega$ are larger than 1 (recall that we are actually suggesting the values $\alpha=1.77$, $\omega=1.45$), then our quadratic term is bigger (more negative) and thus results in a trajectory more bent than the one provided by (\ref{BiressaF,lin}).

\begin{figure*}
\includegraphics[width=\textwidth]{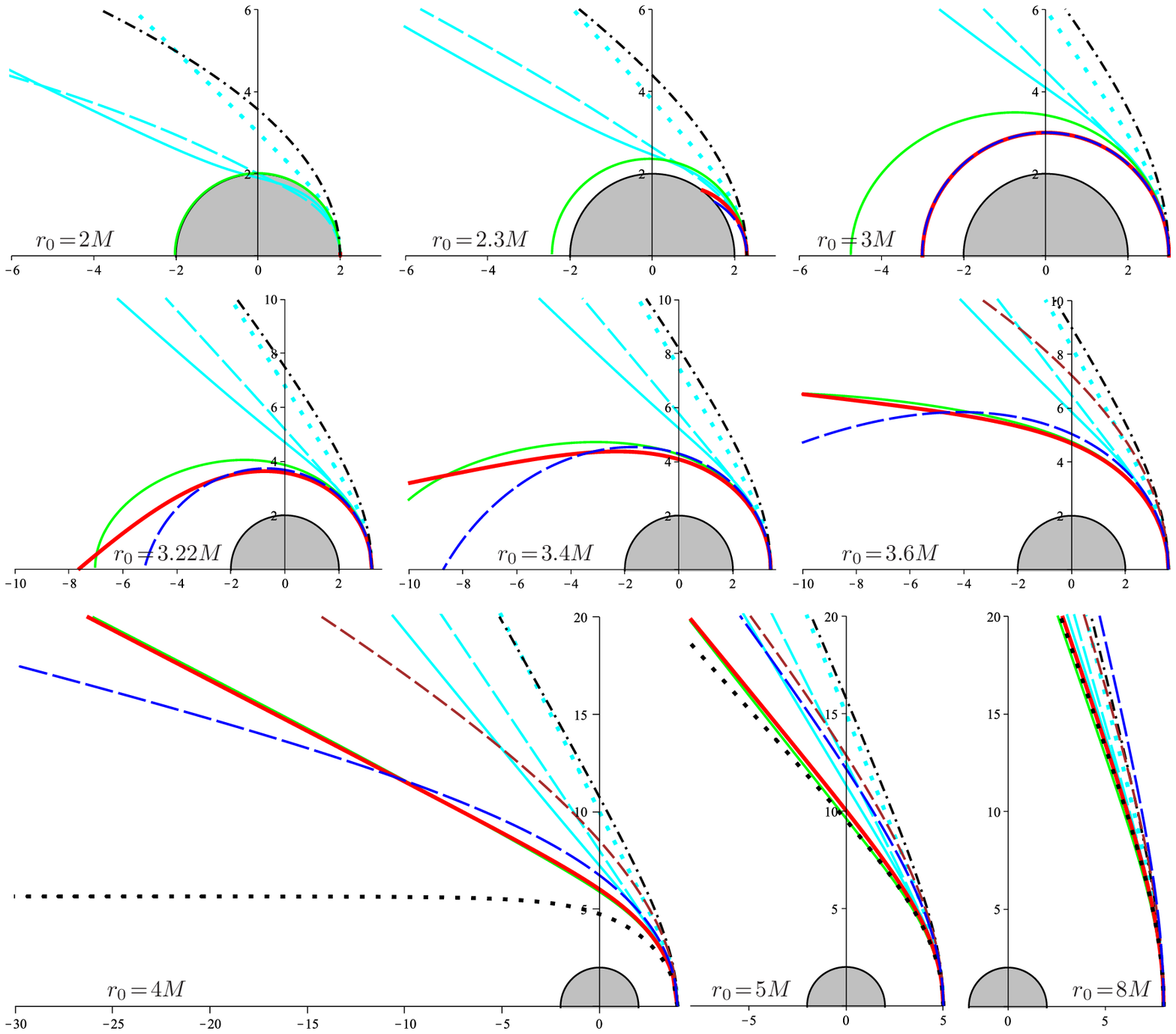}
\caption
{Comparison of the light-ray approximations in the strong field near the black hole. Arrangement and parametrization of the photon trajectories in the equatorial $(r\cos\phi,r\sin\phi)$ plane corresponds to the right plot of figure \ref{ray-description}. The curves show the exact trajectory is in {\it solid red}, the linearization of Darwin's formula (\ref{Darwin,lin}) yields the {\it dot-dashed black} line, the linearized solution of the Binet formula by \cite{BhadraBS-10} (our equation \ref{BhadraBS,lin}) is {\it dotted cyan}, the one by \cite{BiressaF-11} (our equation \ref{BiressaF,lin}) is the {\it dashed cyan} line, the quadratic-order solution following from \cite{ArakidaK-12} (our equation \ref{ArakidaK,lin}) yields the {\it solid cyan} line, the approximation due to Beloborodov \cite{Beloborodov-02} (our equation \ref{Beloborodov-approx}) is the {\it dotted black} line, a hyperbola (\ref{hyperbola}) adjusted to the given pericentre and to correct asymptotics is the {\it dashed brown} line, the approximation using the pseudo-Newtonian potential recommended by \cite{Wegg-12} (our equation \ref{Wegg-trajectory}) is the {\it dashed (dark) blue} line and our newly suggested approximation (\ref{our-approximation}) is the {\it solid light green} line. From top left to bottom right, the plots show trajectories of photons starting tangentially (from $\phi_0=0$) from radii $r_0=2M$, $r_0=2.3M$, $r_0=3M$, $r_0=3.22M$, $r_0=3.4M$, $r_0=3.6M$, $r_0=4M$, $r_0=5M$ and $r_0=8M$. Beloborodov's prescription (dotted black) can only be employed for $r_0\geq 4M$ and the hyperbola-approximation (dashed brown line) for $\phi_\infty<\pi$. See the main text for further commentary.}
\label{comparison-near}
\end{figure*}

\begin{figure*}
\includegraphics[width=\textwidth]{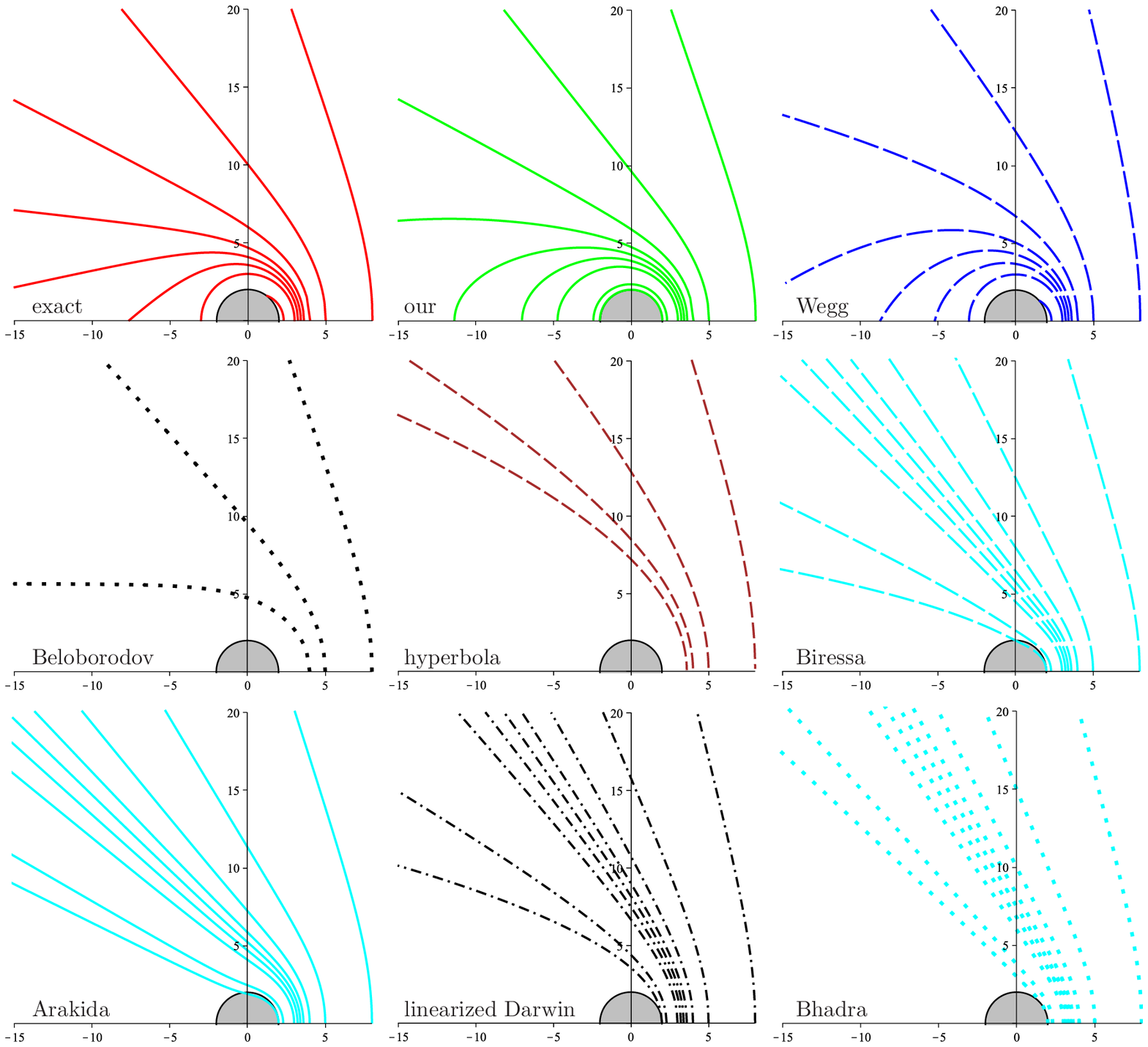}
\caption
{Same curves as those in figure \ref{comparison-near}, i.e. again showing rays with pericentres at $r_0=2M$, $r_0=2.3M$, $r_0=3M$, $r_0=3.22M$, $r_0=3.4M$, $r_0=3.6M$, $r_0=4M$, $r_0=5M$ and $r_0=8M$, but now grouped into plots by approximations (rather than by pericentre radii), so that it is better seen how these depend on radius in comparison with the exact ideal: from top left to bottom right and with the same colouring as in previous figure, one can see exact rays (solid red line), our approximation (solid light green line), pseudo-Newtonian result with Wegg's potential (dashed dark blue line), Beloborodov's approximation (dotted black line; only applicable to the last three trajectories), suitably adjusted hyperbolas (dashed brown line; only applicable to the last four trajectories), the linear approximation by \cite{BiressaF-11} (dashed light blue line), linearized Darwin's formula (dot-dashed black line), the quadratic approximation by \cite{ArakidaK-12} (solid light blue line) and the linear approximation by \cite{BhadraBS-10} (dotted light blue line). The figure reveals that mainly the first-row prescriptions behave satisfactorily down to the very horizon; they are not very accurate down there, but follow the actual rays qualitatively. Also their overall dependence on radius proves to be very close to that visible in the exact pattern.}
\label{near-approximations}
\end{figure*}

\begin{figure*}
\includegraphics[width=\textwidth]{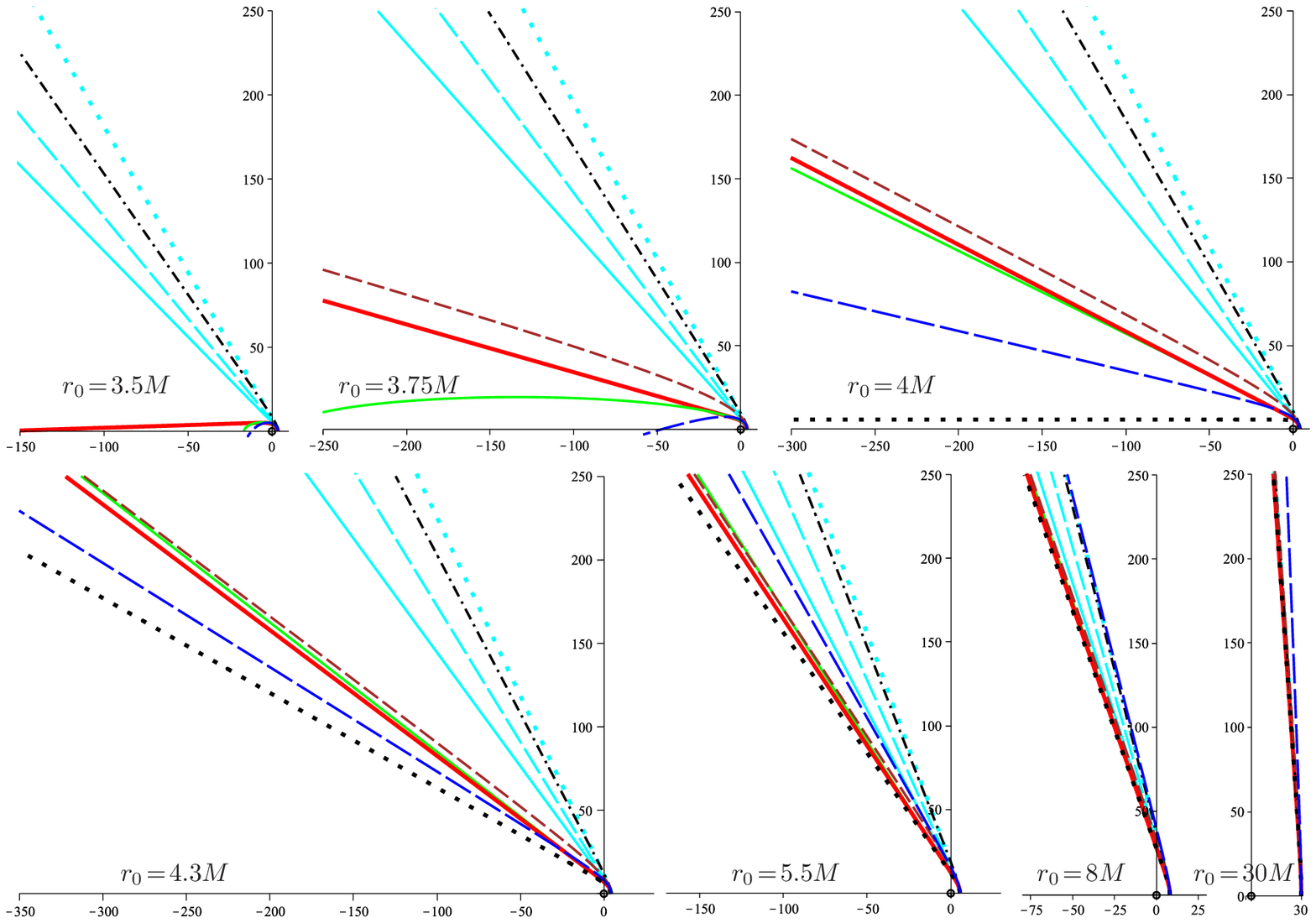}
\caption
{Counterpart of figure \ref{comparison-near} with equatorial $(r\cos\phi,r\sin\phi)$ plots extended to larger radial region, in order to also compare the light-ray approximations in weaker fields farther from the black hole. Again, the exact trajectory is the {\it solid red} line, the linearization of Darwin's formula yields the {\it dot-dashed black} line, the linearized solution of the Binet formula by \cite{BhadraBS-10} is the {\it dotted cyan} line, the one by \cite{BiressaF-11} is the {\it dashed cyan} line, the quadratic-order solution following from \cite{ArakidaK-12} yields the {\it solid cyan} line, the approximation due to \cite{Beloborodov-02} is the {\it dotted black} line, a hyperbola adjusted to the given pericentre and to correct asymptotics is the {\it dashed brown} line, the approximation using the pseudo-Newtonian potential by \cite{Wegg-12} is the {\it dashed (dark) blue} line and our newly suggested approximation is the {\it solid light green} line. From top left to bottom right, the plots show trajectories of photons starting tangentially (from $\phi_0=0$) from radii $r_0=3.5M$, $r_0=3.75M$, $r_0=4M$, $r_0=4.3M$, $r_0=5.5M$, $r_0=8M$ and $r_0=30M$. See the main text for interpretation.}
\label{comparison-far}
\end{figure*}

\section{Comparison on ray examples}
\label{examples}

Let us compare the above approximations of light rays passing at different distances from the centre; we are even including those approaching the horizon very closely. Since the approximate formulas are primarily required to excel at weak-field regions, they cannot be expected to perform well down there. However, it is good to know where and how much they fail. In particular, when using such a formula in a code, it matters whether it is just very inaccurate, or rather yields complete nonsense that has to be discarded.

\begin{figure*}
\includegraphics[width=\textwidth]{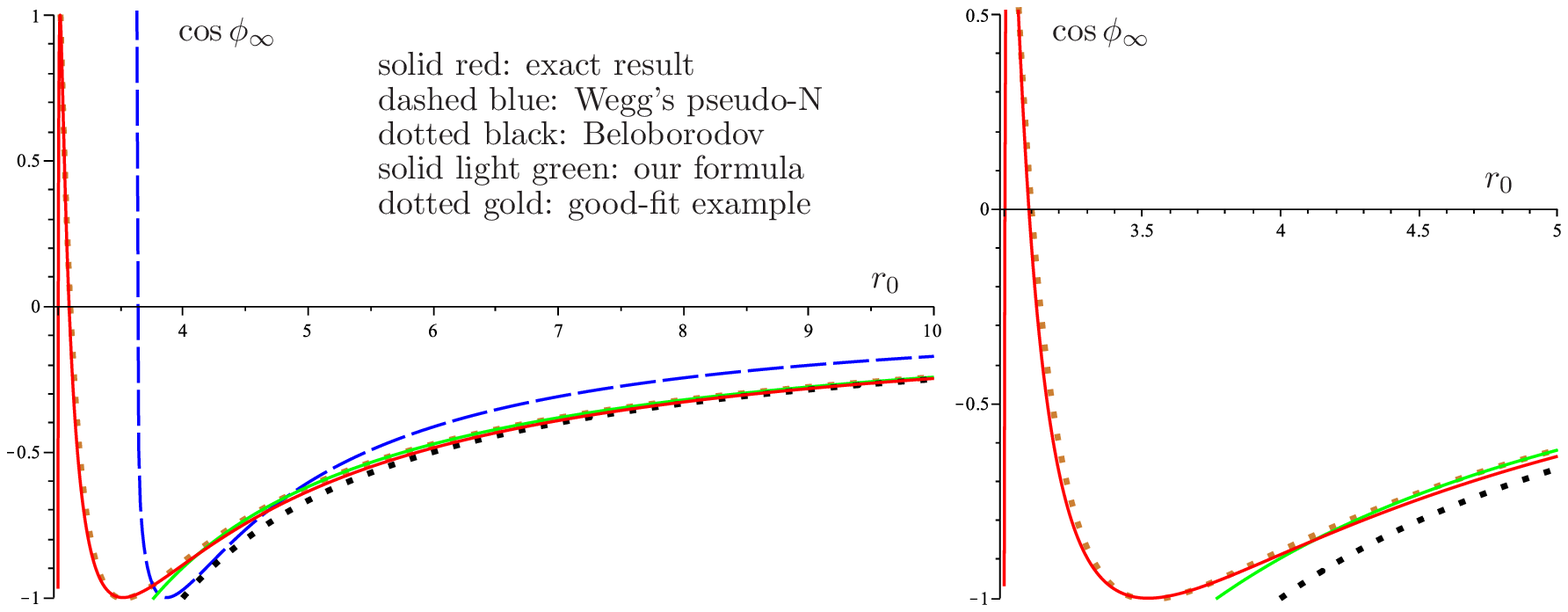}
\caption
{Cosine of the azimuth $\phi$ along which the ray approaches radial infinity, $\cos\phi_\infty$, as given by the exact formula (solid red), pseudo-Newtonian treatment using Wegg's potential (dashed blue), by Beloborodov's approximation (dotted black) and by formula we suggest here (solid green). The right plot provides a more detailed view of the small-pericentre behaviour (the pseudo-Newtonian result is not shown there). Just for the sake of interest, we add a curve (dotted gold) given by the expression (\ref{deflection-formula}) in order to illustrate that it is not difficult to suggest a very good formula solely for the asymptotic angle.}
\label{total-deflection}
\end{figure*}

Numerical results are presented in four figures: figure \ref{comparison-near} compares the approximations at small radii (down to the very horizon), figure \ref{near-approximations} illustrates the radius-dependence of all of them, figure \ref{comparison-far} shows the behaviour at larger radii and \ref{total-deflection} accompanies the remark \ref{asymptotic-angle} on asymptotics (hence total deflection angle) added below.

The figures demonstrate that the approximations obtained by low-order expansions of the exact formulas (namely by linearization or quadratic expansion in $M$) are only usable for rays whose pericentres are above $6M\!\div\!8M$, say. Below this radius, all the ``ad hoc" prescriptions provide much better results, including the pseudo-Newtonian one using the potential suggested by Wegg. Actually, the latter even provides {\em the best} results in a narrow region around the circular photon geodesic at $r=3M$ since it reproduces its location exactly; on the other hand, for pericentres at larger radii it is not as precise as other approximations, though it also falls off to zero deflection correctly. When the pericentre shifts below $4M$, even good approximations become rather problematic. The one by Beloborodov is only usable above this radius, being not very accurate up to some $r=5M$. Our newly suggested formula is very accurate almost everywhere, including the rays with pericentres slightly below $4M$, but at $r_0=3.77M$ it also deviates from the correct behaviour, switching to ``elliptic" behaviour (not at all present in the relativistic treatment, besides the circular photon orbit at $r=3M$) which can only mimic the relativistic trajectories locally. Note that the decisive value $r_0=3.77M$ comes from $r_0=(2+\alpha)M$, so it might be improved (shifted down) by choosing for our constant $\alpha$ a value lower than $1.77$, but this would almost certainly spoil the behaviour farther away.

Ad approximation by a hyperbola: rough as this idea may have seemed, the figures show that it reproduces the large-scale features of the rays very well when tied to a correct asymptotics. However, its bending about the black hole is not sharp enough and, also, just when endowed with {\em correct} asymptotics, it gets quite complicated and not invertible for $r_0$.

Finally, it is important that the two approximations which are satisfactory in general and can be used also below $r_0=4M$, namely our newly suggested formula (\ref{our-approximation}) and the pseudo-Newtonian result using Wegg's potential, serve acceptably even there (at $4M>r_0>2M$), mainly they nowhere return {\em error}. Above $r_0\simeq 5M$, the approximation by Beloborodov remains a benchmark whose main advantage is the extraordinarily simple relation (\ref{alpha-r,psi}) between the angular position on the trajectory and the latter's local direction.

\section{Asymptotic angle}
\label{asymptotic-angle}

Asymptotic behaviour of the approximations studied here is clearly visible in figure \ref{comparison-far}, but let us add a special remark (and figure) comparing the $\phi_\infty$ values. It is of course useless to include the approximation by the hyperbola here, because, where applicable, its asymptotic angle has been {\em prescribed} by the correct (exact) value. Therefore, we are left with the results following from the expansion of the exact formula or approximate solution of the Binet formula, with the pseudo-Newtonian result (\ref{Wegg,phi_infty}) using Wegg's potential, with the asymptotics $\cos\phi_\infty=-2M/(r_0-2M)$ of Beloborodov and with $\cos\phi_\infty=-2M/(r_0-\alpha M)$ following from our formula suggested above. These asymptotic forms can naturally be inverted for $r_0$ easier than the general formulas describing the whole trajectory. Inversion is especially simple for Beloborodov's and our approximations, and also for the linearized solution of Binet's formula by \cite{BiressaF-11} given in equation (\ref{BiressaF,lin}):
\begin{align}
  \frac{r_0}{M} &= -\frac{(2+\cos\phi_\infty)(1-\cos\phi_\infty)}{\cos\phi_\infty}
                   & & {\rm Biressa} \\
                &= -\frac{2-2\cos\phi_\infty}{\cos\phi_\infty}
                   & & {\rm Beloborodov} \\
                &= -\frac{2-\alpha\,\cos\phi_\infty}{\cos\phi_\infty}
                   & & {\rm our~formula} \,.
\end{align}

The $\cos\phi_\infty$ plots given in figure \ref{total-deflection} confirm that the exact behaviour (solid red curve) is best reproduced by our formula (green curve), followed by the formula of Beloborodov (dotted black). Note that we only include there results given by the best of the approximations. In particular, we omit those provided by linearizations in $M$ as well as by the quadratic formula. We can supplement the figure by the $r\rightarrow\infty$ limits of the expansions given in section \ref{expansions}, to obtain, up to second order in $M$,
\begin{align}
  \cos\phi_\infty
  &=-\frac{2M}{r_0}-\frac{M^2}{r_0^2}\left(\frac{15\pi}{8}-2\right)
     && {\rm exact} \\
  &=-\frac{2M}{r_0}-\frac{2M^2}{r_0^2}
     && {\rm Biressa} \\
  &=-\frac{M}{r_0}-\frac{M^2}{r_0^2}\left(\frac{3\pi}{2}+1\right)
     && {\rm Wegg} \\
  &=-\frac{2M}{r_0}-\frac{4M^2}{r_0^2}
     && \hspace{-5mm} {\rm Beloborodov} \\
  &=-\frac{2M}{r_0}-\frac{2\alpha M^2}{r_0^2}
     && \hspace{-5mm} {\rm our~formula} \,.
\end{align}
The linear terms are equal, except that of the Wegg's pseudo-Newtonian formula; and the quadratic-term coefficient of the exact value, $-(15\pi/8-2)\doteq -3.89$, is most closely reproduced by Beloborodov ($-4$) and by our formula ($-3.54$ if choosing $\alpha=1.77$ as in the figures).

Let us add that it is not difficult to bring yet a better proposal {\em solely for the asymptotic angle}:
in figure \ref{total-deflection}, look at the dotted gold curve that follows with a slight modification of our asymptotics,
\begin{equation}  \label{deflection-formula}
  \cos\phi_\infty=-\frac{2M}{r_0-\alpha M}\left[1-\frac{M^6}{(r_0-2.1M)^6}\right].
\end{equation}
It mirrors the correct curve very accurately down to some $r_0=3.02M$ where it reaches $\cos\phi_\infty=+1$, namely the photon makes a $360^\circ$ angular revolution when starting from there.
We stress that this has been just an ad hoc example; without doubt, still more accurate asymptotic formulas can be found.

\begin{figure*}
\epsscale{.8}
\plotone{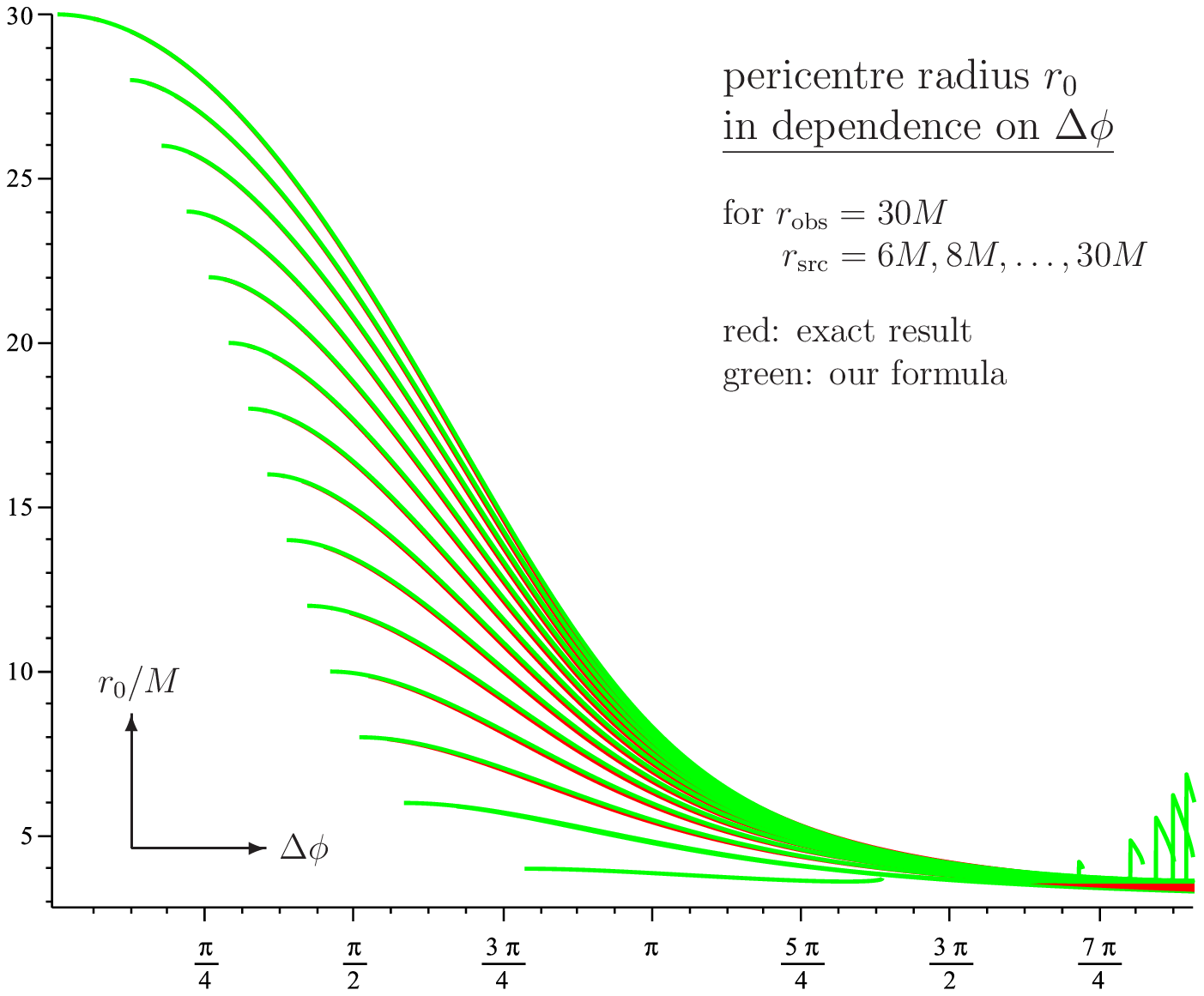}\caption
{Finding the connecting ray for a given configuration of source ($r=r_{\rm src}$) and observer ($r=r_{\rm obs}$), when only their angular separation $\Delta\phi$ is known (not their absolute angular positions). The solution is given in terms of the pericentre radius $r_0$ as a function of $\Delta\phi$, for $r_{\rm obs}=30M$; each curve corresponds to one particular value of $r_{\rm src}$, specifically, (from bottom to top) $r_{\rm src}=6M$, $8M$, $10M$, \dots, $30M$. Green curves have been obtained using our approximation (\ref{our-approximation,r0}), while red curves (drawn ``under'' the green ones) follow by numerical solution from the exact formula. The green approximation is clearly very accurate, it only fails at $\Delta\phi\rightarrow 2\pi$ ($\Leftrightarrow$ bending by $\sim\pi$) where our formula switches to bound-orbit mode (this happens for $r_0<(2+\alpha)M=3.77M$, specifically).
}
\label{exercise}
\end{figure*}

\section{Finding the ray parameters for generic boundary conditions}
\label{lensing-exercise}

We have assumed that the azimuthal angle $\phi$ is adjusted so that $r(\phi=0)=r_0$, but in a real lensing situation one can only suppose to know positions of the source, of the lensing body and of the observer. Placing the coordinate origin at the lensing body and choosing the azimuthal (``equatorial") plane as that defined by the connecting ray, this means knowing the radii of the source and of the observer, plus the angular {\em difference} between the source and the observer, $\Delta\phi$ say. One a priori does not know the angular position of the ray pericentre, so can{\em not} fix the azimuth $\phi$ absolutely prior to solving our inversion exercise $(r,\phi)\rightarrow r_0$. Consider now whether the exercise is still solvable, i.e. whether it is possible to find $r_0$ from this accessible information.\footnote
{I thank the referee for drawing my attention to this point.}

In order for the pericentre $r_0$ to have proper sense, let us assume that it lies between the source and the observer. As we wish to {\em eventually} (in solving the problem) adjust the azimuth $\phi$ so that $r(\phi=0)=r_0$, we can assume (for example) that the source is at $ \phi<0$ and the observer is at $\phi>0$. Let us thus denote their positions by $(r_{\rm src}\!>\!r_0,-\pi<\phi_{\rm src}\!<\!0)$ and $(r_{\rm obs}\!>\!r_0,\pi>\phi_{\rm obs}\!>\!0)$, respectively. Imagine solving the inversion for $r_0$ ``from both sides", i.e. looking for $(r_{\rm src},\cos\phi_{\rm src})\rightarrow r_0$ and $(r_{\rm obs},\cos\phi_{\rm obs})\rightarrow r_0$ (regarding that $\phi_{\rm src}\!<\!0$ whereas $\phi_{\rm obs}\!>\!0$, it is better to write the angular data in terms of the cosine which is independent of the sign; the exercise can then be treated using the same formulas ``from both sides" of the $\phi=0$ plane). Both must lead to the same pericentre $r_0$, and we also know that $\phi_{\rm obs}\!-\!\phi_{\rm src}$ gives the total angular distance travelled, $\Delta\phi$ (supposed to be $<2\pi$), so we have two constraints
\begin{equation}
  r_0(r_{\rm src},\phi_{\rm src})=r_0(r_{\rm obs},\phi_{\rm obs}),
  \quad
  \phi_{\rm obs}\!-\!\phi_{\rm src}=\Delta\phi.
\end{equation}
Since $r_{\rm src}$, $r_{\rm obs}$ and $\Delta\phi$ are known, we have two equations for two unknowns, $\phi_{\rm src}$ and $\phi_{\rm obs}$.

Let us check whether one can really solve the exercise in such a way. Suppose that we employ our approximation according to which $r_0$ is given in terms of $r$ and $\cos\phi$ by equation (\ref{our-approximation,r0}). The constraints thus yield
\begin{equation}
  \frac{\left({\cal R}+\sqrt{{\cal R}^2+4Mr{\cal A}{\cal B}}\right)_{\rm src}}
       {2{\cal A}_{\rm src}}
  =
  \frac{\left({\cal R}+\sqrt{{\cal R}^2+4Mr{\cal A}{\cal B}}\right)_{\rm obs}}
       {2{\cal A}_{\rm obs}},
\end{equation}
where
\begin{align*}
  {\cal R}_{\rm src} &:= {\cal R}(r\!=\!r_{\rm src},\phi\!=\!\phi_{\rm src}), \\
  {\cal R}_{\rm obs} &:= {\cal R}(r\!=\!r_{\rm obs},\phi\!=\!\phi_{\rm src}\!+\!\Delta\phi), \\
  {\cal A}_{\rm src,obs} &:= {\cal A}(r\!=\!r_{\rm src,obs}), \\
  {\cal B}_{\rm src} &:= {\cal B}(r\!=\!r_{\rm src},\phi\!=\!\phi_{\rm src}), \\
  {\cal B}_{\rm obs} &:= {\cal B}(r\!=\!r_{\rm obs},\phi\!=\!\phi_{\rm src}\!+\!\Delta\phi),
\end{align*}
with ${\cal R}$, ${\cal A}$, ${\cal B}$ introduced below equation (\ref{our-approximation,r0}).
The above equation is to be solved for $\phi_{\rm src}$ which in turn implies $\phi_{\rm obs}(=\!\phi_{\rm src}\!+\!\Delta\phi)$ and $r_0$, all as functions of $r_{\rm src}$, $r_{\rm obs}$ and $\Delta\phi$.

As an illustration (figure \ref{exercise}), let us choose $r_{\rm obs}\geq r_{\rm src}$ (without loss of generality) and monitor how the solution of the exercise changes with $\Delta\phi$ increasing from zero to $2\pi$ for a series of source radii $r_{\rm src}$ increasing from some small value to $r_{\rm obs}$. A simple chart with the source lying somewhere on the $r=r_{\rm src}$, $-\pi<\phi<0$ half-circle, the observer lying on the $r=r_{\rm obs}\geq r_{\rm src}$, $\pi>\phi>0$ half-circle, and their angular separation $\Delta\phi$ gradually growing, reveals that
i) when $r_{\rm src}$ and $r_{\rm obs}$ are sufficiently different, the solution does not exist for too-small $\Delta\phi$ (the connecting ray is nowhere purely tangential to the centre then);
ii) the solution only starts to exist when $\Delta\phi$ is large enough for $r_{\rm src}$ to just coincide with $r_0$;
iii) within the interval of $\Delta\phi$ where the solution does exist, the pericentre $r_0$ typically decreases from $r_{\rm src}$ with increasing $\Delta\phi$, because increasing the angular separation corresponds to a connecting ray increasingly bent around the centre;
iv) for $\Delta\phi\rightarrow 2\pi$ the pericentre radius falls almost to $3M$ and the approximations more or less cease to provide reasonable answers; specifically, we learned at the end of section \ref{new-suggestion} that according to our approximation the minimal possible pericentre of an unbound trajectory lies at $(2+\alpha)M=3.77M$.

The above estimates are confirmed by figure \ref{exercise} where the pericentre radii are plotted, in dependence on $\Delta\phi$, for $r_{\rm obs}=30M$ and $r_{\rm src}=6M$, $8M$, $10M$, \dots, $30M$.
Besides the curves $r_0(r_{\rm obs},r_{\rm src};\Delta\phi)$ obtained from our approximation (green curves), the figure also shows analogous results obtained, purely numerically, from the {\em exact} formula (red curves, drawn ``under" the green ones). Clearly the two series of curves almost coincide, even at quite small $r_0$; our approximation is only not usable for very large $\Delta\phi$ (approaching $2\pi$), because this corresponds to very large bending (by $\sim\pi$) where our formula goes over to the elliptic-type bound-orbit regime.

\section{Concluding remarks}

We have suggested a new formula approximating light rays in the Schwarzschild space, compared it with other formulas from the literature and showed that it performs very well, though being quite simple and easily invertible for the pericentre radius $r_0$. Besides analytical estimates, the main focus has been in numerical testing of various plausible approximations against exact results in a very strong as well as a weaker-field regime. We have shown that our formula also yields very good results for the asymptotic angle of photons, as well as in searching for a connecting ray in a given source--gravitating body--observer configuration.

On a more general level, we can conclude that in spite of the legitimate vigilance toward ad hoc prescriptions, not following from an exact result by any sound procedure, in our comparison such formulas proved considerably better than low-order expansions of the exact formula, some of them providing acceptable results even in close vicinity to the horizon.

\acknowledgments
I am thankful for support from Czech grant GACR-14-10625S.

\end{document}